\documentclass[a4paper,twocolumn,showpacs]{revtex4}
\usepackage{amsmath}
\usepackage{graphicx}
\usepackage[english]{babel} 
\usepackage[latin1]{inputenc} 
\usepackage[T1]{fontenc}

\begin{document}

\title{A general creation-annihilation model with absorbing states}
\author{Wellington G. \surname{Dantas}}
\email{wgd@if.uff.br}  
\author{Armando \surname{Ticona}}
\altaffiliation[Present address: ] {Instituto de Investigaciones Fisicas\\
Universidad Mayor de San Andres\\
Casilla 8635, La Paz, Bolivia}
\author{J\"urgen F. \surname{Stilck}}
\email{jstilck@if.uff.br}  
\affiliation{Instituto de F\'{\i}sica, Universidade Federal 
Fluminense, \\
Campus da Praia Vermelha,  \\
Niter\'oi, RJ, 24.210-340, Brazil.}

\date{\today}

\begin{abstract}
A one dimensional non-equilibrium stochastic model is proposed where each site
of the lattice is  occupied by a particle, which may be of type A or
B. The time evolution of the model occurs through three processes:
autocatalytic generation of A and B particles and spontaneous conversion A
$\to$ B. The two-parameter phase diagram of the model is obtained in one- and
two-site mean field approximations, as well as through numerical
simulations and exact solution of finite systems extrapolated to the
thermodynamic limit. A continuous line of transitions between an active and an
absorbing phase is found. This critical line starts at a point where the model
is equivalent to the contact process and ends at a point which corresponds to
the voter model, where two absorbing states coexist. Thus, the critical line
ends at a point where the transition is discontinuous.  Estimates of critical
exponents are obtained through the simulations and finite-size-scaling
extrapolations, and the crossover between universality classes as the voter
model transition is approached is studied. 
\end{abstract}


\maketitle

\section{Introduction}

\label{intro}

The phase transitions exhibited by stochastic models with absorbing states
have attracted much attention in recent years, particularly in order to
identify and understand the aspects which determine the universality classes
in those models. Most of these models have not been solved exactly, but a
variety of approximations allow quite conclusive results regarding their
critical properties. Stochastic models are, of course, well fitted for
simulations, but closed form approximations and other analytical approaches
have also been useful in investigating their behavior \cite{md99}.

One of the simplest and most studied model of this type is the contact process
(CP), which was conceived as a simple model for the spreading of an
epidemic and proven to display a continuous transition between the absorbing
and an active state, even in one dimension \cite{h74}. Actually, it was found
that the CP is equivalent to other models such as Schl\"ogl's lattice model
for autocatalytic chemical reactions \cite{s72} and Reggeon Field Theory (RFT)
\cite{gt}. The CP belongs to the direct percolation (DP) universality class,
together with others models such as the Ziff-Gulari-Barshad model of catalysis
\cite{zgb86} and branching and annihilating walks with an odd offspring
\cite{tty92}. The {\em DP conjecture} states that all phase transitions between
an active and an absorbing state in models with a scalar order parameter,
short range interactions and no conservation laws belong to this class
\cite{j81}. This conjecture was verified in all cases studied so far
\cite{h00}.

Here we study a generalization of the CP, with an additional parameter, 
so that the CP transition point becomes a critical line. This model is similar
to the model proposed by Hinrichsen \cite{h97} in the particular case where
his parameter $q$ is set equal to one. Since the symmetry
properties of this generalized model are the same of the CP, it is expected
that this critical line should belong to the DP universality class. However,
at one point of this line the model is equivalent to the zero temperature
Glauber model \cite{g63}, also called the voter model \cite{l85}, which
displays a spin 
inversion (or particle-hole) symmetry and therefore belongs to another
universality class (the compact directed percolation (CDP) class). Thus the
critical line in the phase diagram of the 
generalized model starts at the CP model and ends at the voter model, a
crossover between the two universality classes being observed.

In section \ref{mod} we define the model and show its equivalence to the CP
and the voter model in the appropriate limits. The phase diagram of the model
is obtained in one- and two-site approximations in section \ref{mf}. In
section \ref{mc}, results of simulations are shown which lead to numerical
estimates of the critical line and of dynamic critical exponents. Other
estimates of the critical line and static critical exponents are obtained
through the exact diagonalization of the time evolution operator of the model 
for finite lattices, extrapolated to the thermodynamic limit using finite size
scaling in section \ref{ed}. Final comments and the conclusion may be found in
section \ref{con}.

\section{Definition of the model}
\label{mod}

Each one of the $N$ sites of a one-dimensional lattice with periodic boundary 
conditions are occupied by particles A or B. No holes are allowed.  Thus, the 
state of the system at a given time $t$ is described by the occupation 
variables $\eta =(\eta_1,\eta_2,...,\eta_N)$, where $\eta_i=0$ or $1$ if site
$i$ is  occupied by particles B or A, respectively.

The time evolution of the system is defined by the following Markovian rules:

\begin{enumerate}

\item A site $i$ of the lattice is chosen randomly.
\item If the site is occupied by a particle B, it becomes occupied with 
a particle A with a transition rate
equal to $p_a n_a/2$, where $n_a$ is the number of A particles in 
first neighbor sites of $i$.

\begin{figure}[h!]
\begin{center}
\includegraphics[height=1.2cm]{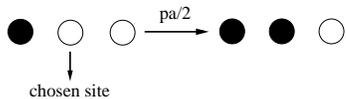}
\caption{Autocatalytic creation of a particle A at site $i$. Full circles
represent A particles and empty circles denote B particles.}
\label{ex1}
\end{center}
\end{figure}

\item If site $i$ is occupied by a particle A, it may become a 
particle B through two processes 
\begin{itemize}
\item Spontaneously, with a transition rate $p_c$

\begin{figure}[h!]
\begin{center}
\includegraphics[height=1.2cm]{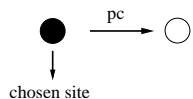}
\caption{Example of spontaneous creation of particle B.}
\label{ex2}
\end{center}
\end{figure}
\item Through an autocatalytic reaction, with a rate $p_bn_b/2$, where
$n_b$ is the number of B particles in the first neighbors of site $i$.

\begin{figure}[h!]
\begin{center}
\vspace{0.3cm}
\hspace{0.8cm}\includegraphics[height=1.2cm]{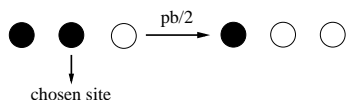}
\caption{Example of autocatalytic creation of a particle B.}
\label{ex3}
\end{center}
\end{figure}
\end{itemize}
\end{enumerate}
\vspace*{-0.6cm}
We define the time step in such a way that the parameters $p_a$, 
$p_b$, and $p_c$, which are non-negative, obey the normalization 
$p_a+p_b+p_c=1$, so that
only two of them are independent. For convenience, we will discuss the
behavior of the model in the $(p_a,p_c)$ plane without loss of 
generality.

The probability $P(\eta,t)$ to find the system in state $\eta$ at 
time $t$ obeys the master equation
\begin{eqnarray}
\label{eq1}
\frac{\partial P(\eta,t)}{\partial t} = 
\sum_m\{w_i(\eta^i)P(\eta^i,t),
-w_i(\eta)P(\eta,t)\}
\end{eqnarray}
where $\eta^i$ corresponds to the following configuration
\begin{eqnarray}
\label{eq2}
\eta^i\equiv(\eta_1,...,1-\eta_i,...,\eta_N)
\end{eqnarray}
and $w_i(\eta)$ is the transition rate of the model, given by
\begin{eqnarray}
\label{eq3}
w_{i}(\eta) = 
\frac{\lambda}{2}(1-\gamma\eta_i)\sum_{\delta}
\eta_{i+\delta}+\eta_i,
\end{eqnarray}
where $\lambda = p_a/(1-p_a)$, $\gamma = (1-p_c)/p_a$, and the sum is 
over first neighbors of site $i$.  

The equation for the time evolution of the mean number of A particles 
at site $i$, $\langle\eta_i\rangle$, may be obtained from equations 
(\ref{eq1}) and (\ref{eq3}), being given by
\begin{eqnarray}
\label{eq4}
\frac{d\langle\eta_i\rangle}{dt}=\frac{\lambda}{2}\sum_{\delta}\langle\eta_
{i+\delta}[1-(2-\gamma)\eta_i]\rangle-\langle\eta_i\rangle
\end{eqnarray}
and a homogeneous solution, that is, $\rho\equiv\langle\eta_i\rangle$
and $\phi\equiv\langle\eta_i\eta_{i+\delta}\rangle\mbox{  } \forall i$, 
is given by equation
\begin{eqnarray}
\label{eq5}
\frac{d\rho}{dt}=\lambda[\rho-(2-\gamma)\phi]-\rho.
\end{eqnarray}

In principle, we are not able to solve equation (\ref{eq5}) because 
the function $\phi(t)$ is not known. We may write a differential 
equation for $\phi(t)$, but in this equation three variable terms 
such as $\langle\eta_i\eta_j\eta_k\rangle$ will appear, so that a 
infinite hierarchy of equations will be obtained. A systematic 
approximate solution is given below. Although it does not provide 
precise results, it still furnishes a qualitative picture of the 
behavior of the system. 

This model, similarly to what happens in other out of equilibrium 
systems, displays {\em absorbing} states, which are such that once 
they are reached, they will never be left. The evolution rules of the model 
define the state where all sites are occupied by B particles 
($\rho=0$) as absorbing. Besides this stationary state, others may 
exist such that $\rho =\lim_{t \to \infty}\rho(t)$ is nonzero. 
Such states are called {\em active}.

It may be useful to remark that this model may be mapped to a spin 
system if we describe sites occupied by A and B particles by an Ising 
spin variables $\sigma_i=1$ and $\sigma_i=-1$, respectively. 
In these variables, 
the transition rate will be given by
\begin{eqnarray}
\label{eq6}
w_{i}^{\uparrow\downarrow}(\sigma) = 
\frac{\alpha}{2}\left[1+\beta \sigma_i-\frac{1}{2}(\epsilon 
\sigma_i+\xi)
\sum_{\delta}\sigma_{i+\delta}\right],
\end{eqnarray}
where $\alpha =(p_a+p_b+2p_c)/2,\beta=(p_a-p_b-2p_c)/(pa+p_b+2p_c),
\epsilon=(p_a+p_b)/(p_a+p_b+2p_c)$, and  
$\xi=(p_a-p_b)/(p_a+p_b+2p_c)$.

Finally, this model corresponds to two known models in particular 
limits. If we make $p_b=0$ or $\gamma=1$ the well known contact 
process is recovered \cite{h74},
\begin{eqnarray}
\label{eq7}
w_{i}^{(CP)}(\eta)=\frac{\lambda}{2}(1-\eta_i)\sum_{\delta}
\eta_{i+\delta}+\eta_i.
\end{eqnarray}
If now we take $p_a=p_b$ and $p_c=0$ in the spin formulation of the 
model, the zero temperature linear Glauber model is recovered, also 
known as the voter model,
\begin{eqnarray}
w_{i}^{(LGM)}(\sigma) = \frac{\alpha}{2}\left[1-\frac{1}{2}\sigma_i
\sum_{\delta}\sigma_{i+\delta}\right].
\end{eqnarray}

\section{Mean Field Solution}
\label{mf}

An approximate solution of equation (\ref{eq5}) is obtained if we 
write $\phi$ as a function of $\rho$. This approximation does not 
account for correlations and is known as the one site mean field 
approximation, assuming 
$\langle\eta_i\eta_j\rangle=\langle\eta_i\rangle\langle\eta_j\rangle$, 
that is $\phi=\rho^2$. Within this approximation, equation (\ref{eq5}) 
may be written as
\begin{eqnarray}
\frac{d\rho}{dt}=(\lambda-1)\rho-(2-\gamma)\lambda\rho^2,
\end{eqnarray}
and the stationary solution in terms of the parameters $p_a$ and 
$p_c$ is
\begin{eqnarray}
\rho=\left\{\begin{array}{ll}
0&\ \mbox{if $p_a\le 1/2$}\\
\frac{2p_a-1}{2p_a+p_c-1}&\mbox{otherwise.}
\end{array}\right.
\end{eqnarray}

\begin{figure}[h!]
\begin{center}
\includegraphics[height=6cm]{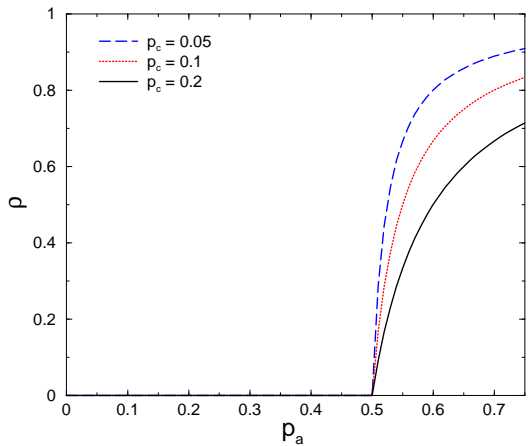}
\caption{Density in the steady state as a function of $p_a$ for some 
fixed values of $p_c$.}
\label{dens1st}
\end{center}
\end{figure}

Therefore, in the one-site approximation the absorbing state is 
separated from the active state by a continuous transition  line 
located at $p_a=1/2$ and the behavior of $\rho$ as a function of 
$p_a$ for some values of $p_c$ is shown in figure (\ref{dens1st}). 
\vspace*{-0.5cm}
As expected, as $p_c$ is increased, the density of A sites decreases, 
since annihilation of A particles is favored. It may be shown that 
close to the transition line the order parameter behaves as
$\rho\sim\Delta^{\beta}$,  $\Delta=(p_a-1/2) > 0$, with 
$\beta=1$ and, at the critical line, the density decays as
$\rho\sim t^{-\varphi}$, with $\varphi=1$. These exponents are identical to 
the mean field results for the contact process.

A somewhat better approximation is obtained if the set of equations 
for $\rho$ and $\phi$ is simultaneously solved. As remarked above, an 
equation for $\phi = \langle\eta_i\eta_j\rangle$ shows mean values of 
products of three $\eta$ variables. Therefore, in order to get a 
closed set of two equations, these mean values should be approximated 
as functions of $\phi$ and $\rho$. In doing so, we are neglecting 
higher order correlations and this is known as the two-site mean 
field approximation.

The following notation is helpful to obtain the second differential 
equation of the two-site approximation. We call $\rho=P(\bullet)$, 
where $P(\bullet)$ is the probability that a site is occupied by an A 
particle and the probability to find a B particle is 
$P(\circ)=1-P(\bullet)$. We also define the three probabilities of 
the configurations of two neighboring sites, and using the 
relations
\begin{eqnarray}
P(\circ)&=&P(\bullet\circ)+P(\circ\circ),\nonumber\\
P(\bullet)&=&P(\bullet\circ)+P(\bullet\bullet),\\
\label{eq8}
\end{eqnarray}
together with the evolution rules stated above, the following 
equation for $P(\bullet\circ)$ is found
\begin{eqnarray}
\label{eq81}
\frac{dP(\circ\bullet)}{dt}&=&\frac{1}{2}\{p_a[P(\circ\circ\bullet)-
P(\bullet\circ\bullet)]\nonumber\\&+&
p_b[P(\circ\bullet\bullet)-P(\circ\bullet\circ)]\}\nonumber\\
&+&p_cP(\bullet\bullet)-\frac{1}{2}(1+p_c)P(\circ\bullet).
\end{eqnarray}
As already mentioned the probabilities of three site clusters appear 
in the equation. These will be written in terms of two-site 
probabilities through the so called pair approximation
\begin{eqnarray}
\label{eqaprox}
P(n_1n_2n_3)\approx\frac{P(n_1n_2)P(n_2n_3)}{P(n_2)}.
\end{eqnarray}

Within this approximation, a closed system of two equations is found
\begin{eqnarray}
\label{eq9}
\frac{d\rho}{dt}&=&(p_a-p_b)u-p_c\rho\\
\label{eq10}
\frac{du}{dt}&=&\frac{1}{2}\left[p_a\frac{ru-u^2}{1-\rho}+
p_b\frac{su-u^2}{\rho}\right]\nonumber\\
&+&p_c s-\frac{1}{2}(1+p_c)u,
\end{eqnarray}
where $u\equiv P(\bullet\circ), r\equiv P(\circ\circ)=1-\rho-u$ and
$s\equiv P(\bullet\bullet)=\rho-u$. The first equation above 
corresponds to equation (\ref{eq5}), whereas the second is obtained 
from equation (\ref{eq81}) using the pair approximation 
(\ref{eqaprox}).

\begin{figure}[h!]
\begin{center}
\includegraphics[height=6cm]{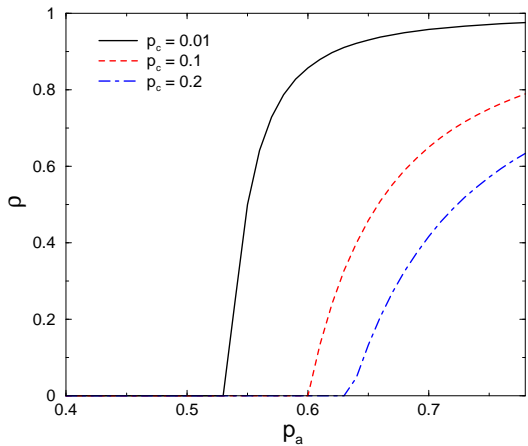}
\caption{Density as a function of $p_a$ for some values of $p_c$, results of
the two-site approximation.}
\label{den2st}
\end{center}
\end{figure}

The stationary solutions for the densities are
\vspace*{-0.3cm}
\begin{eqnarray}
\rho = \left\{\begin{array}{ll}
0 & \textrm{ if  $p_a < p_a^c$}\\
\\
\frac{4p_a^2-4p_a+p_ap_c+1-p_c}{4p_a^2-4p_a+2p_ap_c+1-p_c}& 
\textrm{otherwise},
\end{array}\right.
\end{eqnarray}
where $p_a^c=\frac{1}{8}(4-p_c+\sqrt{8p_c+p_c^2}) $. In this approximation 
the critical value of $p_a$ is a function of $p_c$. The critical 
exponents, as expected, have still their mean-field values, only 
non-universal parameters are different in the two-site approximation 
as compared to the one-site calculation above. Figure (\ref{den2st}) 
shows $\rho$ as a function of $p_a$ for some values of $p_c$ in the 
two-site approximation.

It may be remarked that in the limit $p_c \to 0$ both approximations 
show a discontinuous transition between two absorbing states, which 
happens at $p_a=1/2$. In this limit the model corresponds to a spin 
model where a given spin is reversed with a probability proportional 
to the number of spins in first neighbor sites which are in the 
opposite direction. This might be called a biased voter model when 
$p_a$ is different from $p_b$.

\section{Simulations}
\label{mc}

Although, as seen above, the model exhibits an active state in part of 
the parameter space in the approximate solutions, numerical 
simulations are useful providing some confidence that this state is 
not an artifact of the approximation. To our knowledge, for 
non-equilibrium models an argument similar to the Peierls construction 
for equilibrium models is still not available, and therefore no 
systematic proof of the existence of the active state may be given, although
this was accomplished using probabilistic arguments by Harris in his original
work on the contact process \cite{h74}.

The simulation is done for discrete time, and may be described by the 
following steps:
\begin{enumerate}
\item Initially, the system of $N$ sites with periodic boundary 
conditions is in a state with just one A particle.
\item A list of all sites occupied by A particles is stored, and at 
each time step one of them is chosen randomly.
\item Once the site is chosen, a random number $p$ uniformly 
distributed in the interval $[0,1]$ is generated, if $p < p_a$ a B 
particle is replaced by an A particle in one of the first neighbors, if 
possible. Otherwise, the A particle at the chosen site will be turned 
into a B particle either through the spontaneous or through the 
autocatalytic process.
\item To define the process switching A to B, another random number 
$q$ is generated. If $q < p_c/(p_b+p_c)$ the change is spontaneous, 
otherwise it will happen with a probability proportional to the 
number of B particles in first neighbors of the chosen site.
\item The time interval associated with the steps above is  $\Delta 
t=1/N_A$, where $N_A$ is the number of the sites occupied by A particles 
before the step. The process is repeated until either a maximum 
time $t_{max}$ is attained or the absorbing state $N_A=0$ 
is reached.
\item Several runs are done and mean values are calculated as a 
function of time.
\end{enumerate}

\begin{figure}[h!]
\begin{center}
\includegraphics[height=6cm]{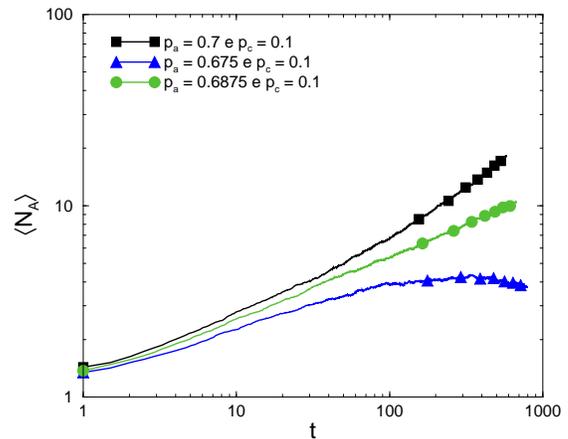}
\caption{Results of simulations }
\label{simu}
\end{center}
\end{figure}

The mean number of A particles, $\langle N_A \rangle$, as a function of time 
for simulations 
with $N=10000$ sites, $t_{max}=100000$ and 10000 repetitions is displayed 
in figure (\ref{simu}). The number of sites is sufficiently high to 
ensure that in the simulations the cluster of A particles is much smaller 
than the system, thus avoiding finite-size effects in the time 
interval considered.
In a region of the $(p_a,p_c)$ plane an active state is found for 
long times in the simulations. Three curves are shown in figure 
(\ref{simu}). The convex one corresponds to an active stationary 
state, while the concave curve signals that the system will reach the 
absorbing state. These two curves are separated by the third, where a 
power law $\langle N_A\rangle\sim t^{\theta}$ is seen, corresponding to the 
critical condition. We adopt, for all critical exponents, the notation 
proposed in the review article by Hinrichsen \cite{h00}.

\begin{figure}[h!]
\begin{center}
\includegraphics[height=6cm]{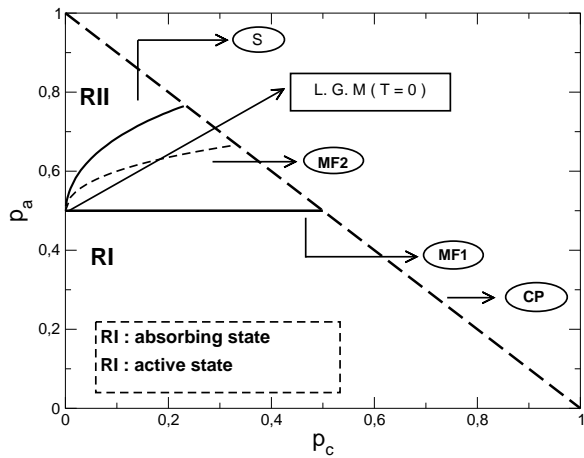}
\caption{Phase diagram of the model, showing estimates for the 
transition line between absorbing and active states in the one (MF1) 
and two (MF2) sites  mean field approximations and simulations (S). 
In the line $p_a+p_c=1$ the model corresponds to the contact process (CP) 
and the unbiased voter model (LGM) is recovered at the point 
$(p_a=1/2,p_c=0)$}
\label{digfases}
\end{center}
\end{figure}

\begin{figure}[h!]
\begin{center}
\includegraphics[height=6cm]{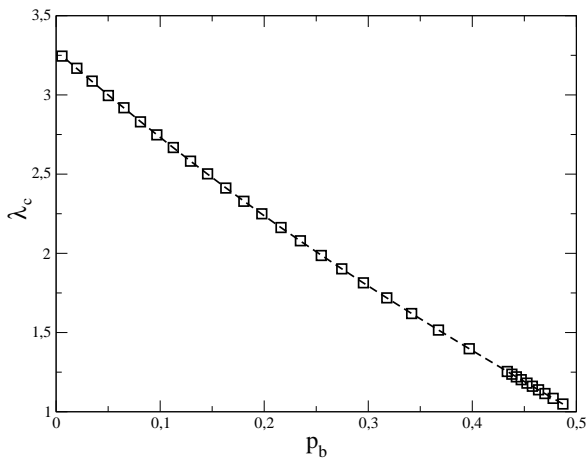}
\caption{Critical values of $\lambda=p_a/(1-p_a)$ as functions of 
$p_b$. The dashed curve is a fit of a fifth degree polynomial to the points
obtained in the simulations.}
\label{lamb}
\end{center}
\end{figure}

Repeating simulations in the region ($p_a+p_c\le 1$), the critical line 
may be estimated. The results of these calculations are shown in 
figure (\ref{digfases}).
The estimate for the critical line provided by the simulations 
display a concavity which agrees with the results of the the two-site 
approximation and, as expected, is situated above the mean field 
results, which usually overestimate the active region in the 
parameter space. The value of $\lambda=p_a/(1-p_a)$ at the critical 
line attains the limiting value $\lambda_c = 3.2945\pm 0.0116$ as
$p_b\to 0$, in good agreement with the estimated critical value for the 
one-dimensional contact process $\lambda_c=3.29785(2)$ \cite{dj91}, 
as may be appreciated in figure (\ref{lamb}).

\begin{figure}[h!]
\begin{center}
\includegraphics[height=6cm]{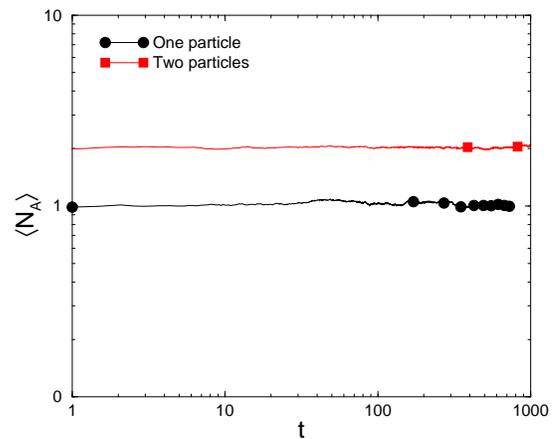}
\caption{Simulational results for the mean number of A particles as a 
function of time for the voter model.}
\label{lambv}
\end{center}
\end{figure}

\begin{figure}[h!]
\begin{center}
\includegraphics[height=6cm]{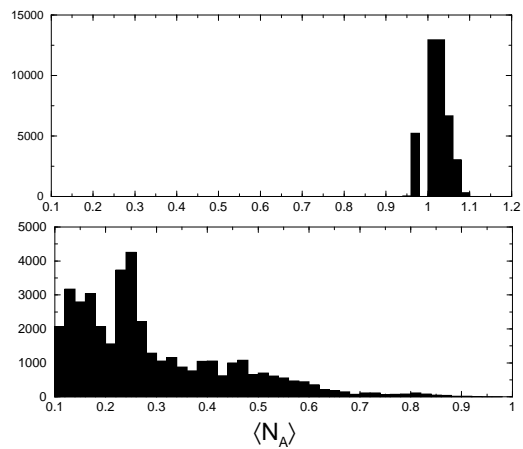}
\caption{Histograms of the mean number of A particles. The upper figure
corresponds to the voter model ($p_a=1/2,p_c=0$) and the other figure is for a
point of  
the parameter space close to the voter model ($p_a=0.49,p_c=0.001$) where the
absorbing state $\langle N_A \rangle=0$ is stable.}
\label{hist}
\end{center}
\end{figure}

The simulational result for $\langle N_A \rangle$ as a function of time in 
the limit 
of the voter model $(p_a=1/2,p_c=0)$ is displayed in figure 
(\ref{lambv}). One notices that the initial number of particles is 
almost conserved, showing a narrow dispersion.
This aspect may also be appreciated in the histograms shown in figure 
(\ref{hist}), for $(p_a=1/2, p_c=0)$  and $(p_a=0.49, p_c=0.001)$, 
with initially one A particle. For the voter model a small dispersion 
is shown around the maximum value $\langle N_A\rangle=1$, while in the 
other case the maximum is shifted to the absorbing state 
$\langle N_A \rangle=0$

The simulations allow us to follow the dynamic evolution of mean values that
describe the model, such as the mean number of A particles $\langle N_A(t)
\rangle$, the probability of survival $P_s(t)$ that at the time $t$ at least
one particle A is present in the system and the mean square radius $R^2(t) =
\langle \sum_i i^2\eta_i(t) \rangle/\langle N_A(t) \rangle$. These variables
satisfy the following scaling forms at the critical point \cite{gt}:
\begin{eqnarray}
\langle N_A \rangle(t)&\sim& t^{\theta},\nonumber\\
P_s(t)&\sim& t^{-\delta},\nonumber\\
R^2(t)&\sim& t^{2/z}.
\end{eqnarray}
Through scaling and hyperscaling relations, these critical exponents define all
other exponents of the model, and thus their values identify the universality
class.

\vspace{0.58cm}
\begin{figure}[h!]
\begin{center}
\includegraphics[height=4.8cm]{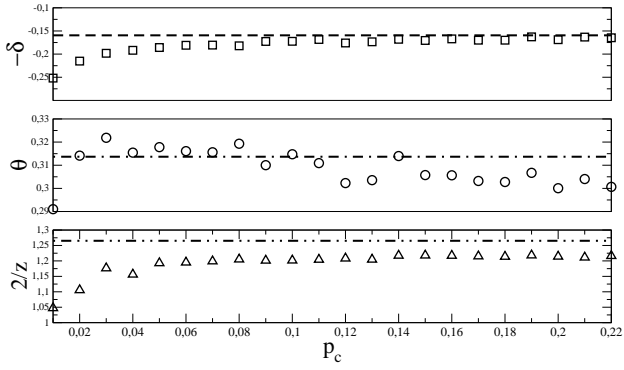}
\caption{Dynamical critical exponents as functions of the rate $p_c$. The
dashed lines indicate the values of these exponents for the CP.}
\label{exps}
\end{center}
\end{figure}
\begin{figure}[h!]
\begin{center}
\includegraphics[height=5cm]{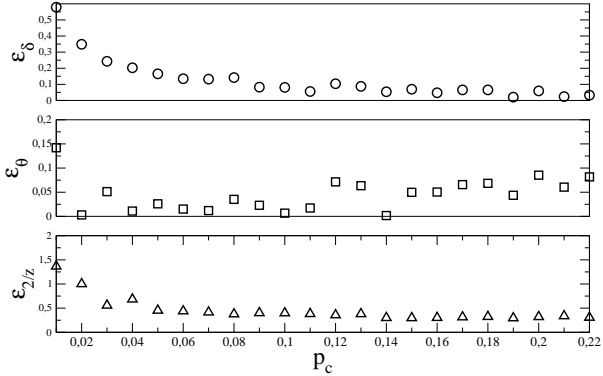}
\caption{Relative errors of the estimates of dynamical critical exponents 
compared to the CP values.}
\vspace{0.1cm}
\label{erros}
\end{center}
\end{figure}
\begin{figure}[h!]
\begin{center}
\includegraphics[height=5cm]{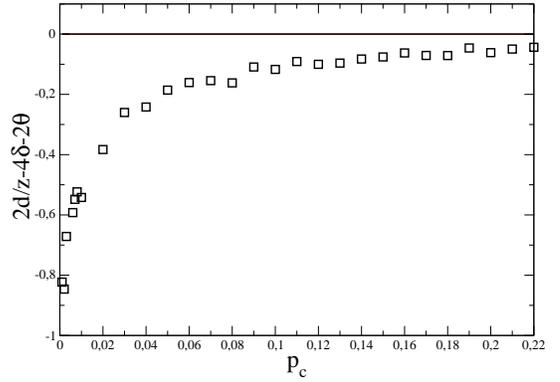}
\caption{Verification of the hyperscaling relation for the dynamical critical
exponents.}
\vspace{0.5cm}
\label{hiper}
\end{center}
\end{figure}

From the simulations, the dynamic critical exponents defined above were
estimated as functions of the rate $p_c$, as is shown in figure (\ref{exps}).
For most of the values for $p_c$, particularly those close to the CP ($p_c
\approx 0.2327$), the estimated exponents are close to the values known for
the CP ($\theta=0.313686$, $\delta=0.159464$, and $2/z=1.265226$
\cite{j99}). The relative errors are 
displayed in figure (\ref{erros}). Also, the hyperscaling
relation $2d/z=4\delta+2\theta$ is satisfied numerically with some
imprecision, as may be seen in figure (\ref{hiper}). Some systematic error
seems to be present in the estimates for 
the mean square radius, which propagates to the hyperscaling relation
verification. As $p_c \to 0$, however, when the voter model is approached, a
departure of all estimates from the CP values is apparent.
\vspace{0.58cm}

\begin{figure}[h!]
\begin{center}
\includegraphics[height=5cm]{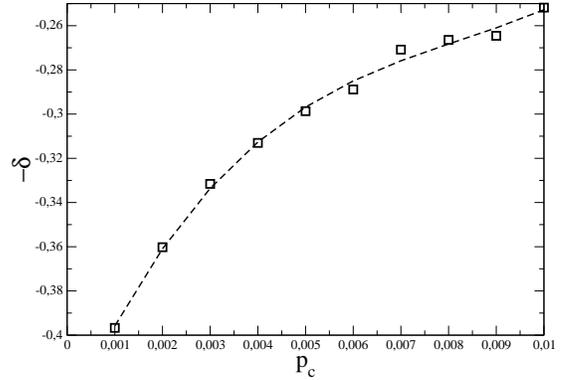}
\caption{Limit for $p_c=0$ of the dynamical exponent $\delta$.}
\label{votaj}
\end{center}
\end{figure}
For models with a discontinuous transition with a spin inversion symmetry,
such as the voter model, the dynamic critical exponents are given by
$\theta_s=0, \delta_s=1/2$ e $2/z_s=1$ and the
hyperscaling relation is given by $\delta_s + \theta_s=d/z_s$, corresponding
to the compact directed percolation (CDP) universality class \cite{h00}. An
analysis of the 
exponents in the limit $p_c \to 0$ may be performed fitting
a cubic curve to the estimates, as is shown in figure (\ref{votaj}) for the
exponent $\delta$. The values obtained through this procedure are
$\delta=0.44$, $\theta=0.07$, and $2/z=1.01$, which are not far from the known
values given above. The variation of the estimates for critical
exponents with the parameters of the model close to $p_c=0$ is
probably an apparent effect, due to large fluctuations observed in the
simulations in this region, and the whole critical line, with
exception of the terminal voter model point, is in the same
universality class of the CP, as suggested by the DP conjecture. This
situation is quite similar to what is observed in the behavior of the
Domany-Kinzel probabilistic cellular automaton (DKPCA) \cite{dk84},
which in a line of its 
phase diagram corresponds to the CP with discrete time and parallel
update, and also exhibits a crossover from the DP to the compact
direct percolation (CDP) universality class \cite{dt95}. It is
believed, although to our knowledge not proven, that changing the
update process in a stochastic model may affect non-universal
quantities only. The point
$p_a=1/2,p_c=0$ may be recognized as a multicritical point, and in its
neighborhood any stationary density variable should exhibit the
scaling form
\begin{equation}
g(p_a-1/2,p_c) \sim (pa-1/2)^{e_g} F\left(\frac{p_c}{[p_a-1/2]^\phi}\right).
\label{sca}
\end{equation}
The critical exponent associated with the density variable $g$, $e_g$,
should correspond to the CDP universality class, and the scaling 
function $F(z)$ is singular at a value $z_0$ of its argument, which
corresponds to the critical line. Thus, the critical line is
asymptotically given by $p_c=z_0(p_a-1/2)^\phi$ and the amplitude 
$z_0$ and the crossover
exponent $\phi$ may be estimated through a fit to the simulational
results. Due to the large fluctuations mentioned above, these estimates are
not very precise. The values we obtained are $z_0=0.47 \pm 0.08$ and
$\phi = 1.80 \pm 0.03$. In the two site approximation, we found $\phi=2$
both for the model described here and for the DKPCA. Although
extensive simulational results for the DKPCA are reported in the
literature \cite{zp94}, apparently no estimate for the crossover
exponent was obtained from them.

\section{Numerical diagonalization}
\label{ed}

Another approach to study 1+1 dimensional stochastic systems is the
exact solution of models with increasing numbers of sites $N$ followed
by extrapolations to the thermodynamic limit \cite{h00}. This is
accomplished writing the master equation (\ref{eq1}) as
\begin{equation}
\frac{\partial }{\partial t} |P(t)\rangle= \hat{\mathcal{S}}|P(t)
\rangle,
\label{me}
\end{equation}
where $\hat{\mathcal{S}}$ is the time evolution operator of the
model. In the representation in which it is diagonal, vanishing
eigenvalues $\mu_0=0$ correspond to stationary states of the
system. For finite systems with absorbing states, only these states are
stationary, and no active stationary state is found.

\begin{figure}[h!]
\begin{center}
\includegraphics[height=5cm]{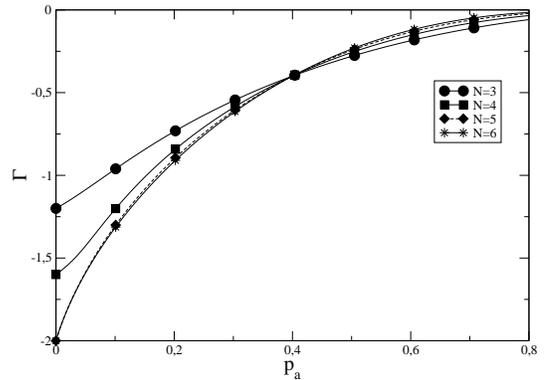}
\caption{Gap $\Gamma$ as a function of $p_a$ for system sizes
$N=3,4,5,6$ with $p_c=0.2$.}
\label{gamma}
\end{center}
\end{figure}

\begin{figure}[h!]
\begin{center}
\includegraphics[height=5cm]{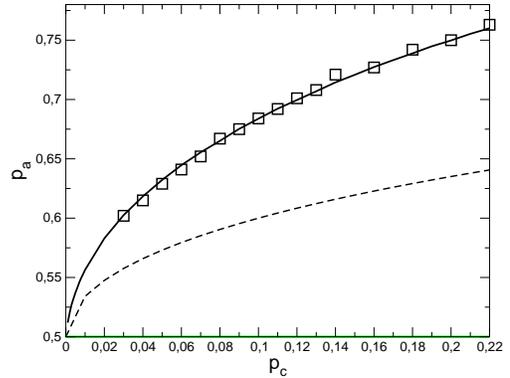}
\caption{The squares are the extrapolated estimates obtained from the
  numerical diagonalization results, compared with the critical lines
  resulting from simulations (full line) and the two-site approximation
  (broken line).}
\label{diag}
\end{center}
\end{figure}

To study the transition between an active and a stationary state, we
may consider the behavior of the eigenvalue with the second smallest
absolute value $\Gamma \equiv \mu_1$ of the operator
$\hat{\mathcal{S}}$. This eigenvalue is related to the
quasi-stationary \cite{dv02} state and will eventually become degenerate with
$\mu_0$ in the thermodynamic limit, originating the phase
transition. Figure (\ref{gamma}) shows the behavior of the gap $\Gamma$ for
some values of the size $N$.
\vspace{0.4cm}
For a fixed value of $p_c$. the finite size scaling
behavior of the gap $\Gamma$ is given by \cite{h00}
\begin{equation}
\Gamma=N^{-z}f(|pa-p_a^c|N^{1/\nu_{\bot}}),
\end{equation}
where $p_a^c$ is the critical value of $p_a$ for a fixed value of
$p_c$. Defining the quantity
\begin{eqnarray}
Y_N(p_a,p_c) = \frac{\ln[\Gamma(p_a,p_c;N+1)/\Gamma(p_a,p_c;N-1)]}
{\ln[(N+1)/(N-1)]},
\end{eqnarray}
we may estimate the critical point $p_a^c(N)$ finding the intersection
of the curves $Y_N$ and $Y_{N+1}$ \cite{chs99}. This procedure
resembles the phenomenological renormalization group. The sequence
$p_a^c(N)$ of estimates for a given value of $p_c$, with $N=4,5,\ldots
M$, was extrapolated to the thermodynamic limit $N \to \infty$ using
the BST algorithm \cite{bst64}, which even for the limited number of
estimates considered lead to rather precise results. Figure (\ref{diag})
\vspace{0.58cm}
shows the extrapolated estimates for the critical line, compared with
the results provided by the simulations and the two-site approximation
results. The agreement between the simulation and the present results
is apparent. The results for the critical line close to the CP point
$p_b=0$ were extrapolated to the CP limit using a cubic function and
the resulting value was $\lambda_c=3.3081\pm 0.0173$, which is in reasonable
agreement with a more precise estimate in the literature
$\lambda_c=3.29785(2)$ \cite{dj91}. We also estimate the crossover 
exponent $\phi$, through a fit to the critical line in the neighborhood of the
voter model limit and it is results in $\phi = 2.24 \pm 0.07$. 

\begin{figure}[h!]
\begin{center}
\includegraphics[height=5cm]{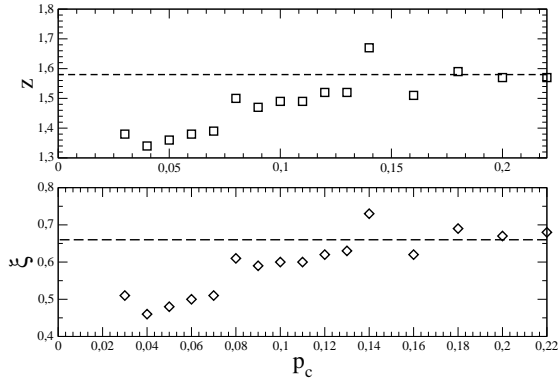}
\caption{Static exponents estimated through the exact diagonalization for
finite systems. The values for the CP are shown in dashed lines.}
\label{exdiag}
\end{center}
\end{figure}

Once the extrapolated estimates for the critical line were obtained,
the critical exponents z and $\xi=z-1/\nu_{\bot}$ may be estimated
using the asymptotic scaling forms 
\begin{eqnarray}
\Gamma(p_a,p_c)&\sim& L^{-z}\nonumber\\
\frac{\partial}{\partial p_a}\Gamma(p_a,p_c)&\sim& L^{-\xi},
\end{eqnarray}
for a fixed value of $p_c$ and $p_a=p_a^c$, on the critical line. An estimate
of the exponents is obtained for each size of the system, and the estimates
are extrapolated to the thermodynamic limit. Our results for these two static
exponents are shown in figure (\ref{exdiag}).
\vspace{0.58cm}

Again a departure of the estimates from the CP values may be observed as $p_c$
becomes smaller. The CDP values for these exponents are $z_s=2$ and
$\xi_s=1$. Unlike to what was observed in the simulations no systematic trend
to these values was observed in the estimates from exact diagonalization as
$p_c \to 0$. We noticed, even considering values of $N$ up to 15, that as
$p_c$ becomes smaller the convergence properties are poorer, and larger sizes
may be necessary to obtain better estimates in this region. The estimates
which were obtained from the simulational data seem to be more reliable that
the ones which follow from the numerical diagonalization results.

\section{Conlusion}
\label{con}

We discussed the transition between an absorbing ($\rho=0$) and an active
($\rho>0$) steady states in a stochastic model of interacting particles. This
model 
displays a line of continuous transition between these states starting at a
point which corresponds to the contact process and ends at a point where the
model is equivalent to the voter model, showing a discontinuous transition
between two absorbing states. The 
phase diagram thus obtained is similar to what is found in a problem of
equilibrium polymerization in a grand-canonical ensemble on an anisotropic
square lattice \cite{s94}, where 
the polymer is modeled as a self-avoiding walk and the activity of a
horizontal link is equal to $x$ and a vertical link is associated with an
activity $y$. A continuous transition is observed in general in the $(x,y)$
plane, between a non-polymerized ($\rho=0$) and a polymerized ($\rho> 0$),
where $\rho$ corresponds to the fraction of lattice sites visited by the
walk. However, at $x=0$ or $y=0$ the walk is one-dimensional, and a
discontinuous transition is found between the non-polymerized and the fully
polymerized ($\rho=1$) phases \cite{pw83}.

The localization of the critical line in the parameter space of the model was
found in the one- and two-site approximations, as well as through simulations
and numerical diagonalization of the time evolution operator. The evidences
obtained from estimates of critical exponents suggest that the whole critical
line belongs to the DP universality class, and a crossover to the CDP
universality class is observed at the terminal voter model point. From the
estimates for the location of the critical line close to the voter model point
the crossover exponent $\phi=1.80 \pm 0.03$ was obtained. Although the
estimated value for $\phi$ is smaller than the mean-field result $\phi=2$, the
latter is inside the confidence interval of the estimate, so that it is not
possible to decide if the crossover exponent has a non-classical value. Due to
the difficulties we had with the simulations and the numerical diagonalization
in the multicritical region, we suspect that another approach is necessary to
clear this point. We are presently addressing this point through a series
expansion approach.
Another extension of this work would be the study of the
crossover between the universality class in the Domany-Kinzel cellular
automaton, which corresponds to the CP with parallel update in a subspace of
its parameter 
space and also has a point of its phase diagram which corresponds to the voter
model.

\begin{acknowledgments}
We thank Prof. Ronald Dickman for many helpful discussions and a critical 
reading of the manuscript. This
research was partially supported by the Brazilian agencies CAPES, CNPq
and FAPERJ, particularly through the project PRONEX-CNPq-FAPERJ/171.168-2003.
\end{acknowledgments}

\end{document}